\documentclass[%
 reprint,
 amsmath,amssymb,
 aps,
]{revtex4-2}

\usepackage{xcolor}
\usepackage{color}
\usepackage{bm}
\usepackage{amssymb, amsmath}
\usepackage{graphicx}
\usepackage{dcolumn}
\usepackage{bm}
\usepackage{hyperref}

\usepackage{soul}

\begin{document}

\preprint{APS/123-QED}

\title{Plasma Turbulence Driven by Wave-Hole Interaction}

\author{Gabriele Celebre}
\email{gabriele.celebre@unical.it}
\author{Mario Imbrogno}
\email{mario.imbrogno@unical.it}
\author{Sergio Servidio}
\author{Francesco Valentini}

\affiliation{Department of Physics, Universit\`a della Calabria, Arcavacata di Rende, 87036, Italy}



\begin{abstract}
Wave–obstacle diffraction is, par excellence, an example of transition to nonlinearity, generating turbulence and complexity in fluids. We present an idealized kinetic plasma regime capturing this ubiquitous interaction and its transition to phase-space turbulence. High-resolution Vlasov–Poisson simulations reveal that the interplay between electrostatic waves and density gaps at Debye and sub-Debye scales redistributes a strongly anisotropic energy cascade throughout the full phase space, unveiling the effect of inhomogeneities on structure formation and small-scale-directed turbulent flow.
\end{abstract}

\maketitle


\textit{Introduction}---In turbulent plasmas, nonlinear wave-like oscillations can drive an energy cascade from large to progressively smaller scales, ultimately leading to dissipation and particle heating \cite{schekochihin2007kinetic,howes2017prospectus}. In weakly collisional systems, this cascade extends deep into the electron kinetic regime, where electromagnetic energy is converted into particle energization \cite{schekochihin2008gyrokinetic,chen2019evidence}. At these fine scales, turbulent dynamics become predominantly electrostatic, with electric-field fluctuations providing an efficient channel for energy transfer to particles \cite{valentini2008cross,camporeale2011dissipation}.

Plasmas are inherently inhomogeneous media, and waves propagate through morphologically complex environments, where low-energy charged particles can become trapped, as described in the electrostatic limit by Bernstein–Greene–Kruskal theory \cite{bernstein1957exact}. Such trapping gives rise to the formation of Debye-scale density holes, whose impact on phase-space dynamics has been extensively investigated \cite{schamel1986electron,hutchinson2017electron,hutchinson2024kinetic}. Their evolution and merging have been explored in laboratory experiments \cite{saeki1979formation}, while analogous structures have been detected by several space missions \cite{ergun1998fast,matsumoto1994electrostatic,kojima1997geotail,franz1998polar}. These coherent structures can be envisioned as obstacles: propagating waves interact with them and are continuously distorted by large populations of holes \cite{cerri2017reconnection}, thereby locally modifying the turbulent cascade. The details of such basic, fundamental interaction remain poorly understood.

In this work, we numerically investigate the kinetic dynamics of high-frequency electrostatic waves propagating through an inhomogeneous electron plasma using a novel 2D–2V time-splitting Vlasov-Poisson (VP) solver. An equilibrium particle distribution function (PDF) for electrons, characterized by density holes, is perturbed by unidirectionally propagating electron-acoustic waves (EAWs), a class of electrostatic fluctuations first identified by \citet{holloway1991undamped} and subsequently reported in heliospheric plasmas \cite{valentini2009electrostatic,vasko2017electron,dillard2018electron,graham2021kinetic} as well as in laboratory experiments \cite{hellberg2000electron,anderegg2009electron,anderegg2009wave,chowdhury2017experimental,affolter2018trapped}.

Owing to their weak Landau damping \cite{holloway1991undamped,valentini2006excitation,valentini2012undamped}, EAWs can survive over long timescales. Enstrophy, i.e., the second Casimir invariant associated with deviations from the equilibrium PDF \cite{knorr1977time,diamond2010modern,servidio2017magnetospheric,nastac2024phase,nastac2025universal}, initially cascades in the 1D–1V phase subspace defined by the EAW propagation direction. Nevertheless, since the local EAW phase velocity vanishes where the density drops to zero, these cavities behave as physical obstacles (analogous to a rock in a river) that scatter the injected, strongly anisotropic one-dimensional turbulence, leading to a full 2D–2V phase-space enstrophy cascade. In this \textit{Letter}, we thoroughly inspect turbulence properties in both fluid and kinetic frameworks through Fourier-Hermite spectral analysis.

\textit{Numerical setup}---The present study examines electrostatic plasma states with a homogeneous proton background and an equilibrium PDF for electrons featuring a spatial depletion in the form of a radially symmetric density hole, $f_h$. This configuration constitutes a stationary solution of the 2D-2V VP system, written in normalized units (see Sec.~A of the Supplemental Material) as:
\begin{subequations}
\label{VPnew}
\begin{gather}
\dfrac{\partial f_{e}}{\partial t} + \bm{v} \cdot \bm{\nabla} f_{e} - \bm{E} \cdot \bm{\nabla}_{\bm{v}} f_e = 0,\label{VP1new} \\[5pt]
\bm{\nabla} \cdot \bm{E} = 1 - \int_{\mathbb{R}^2} f_e \, d^2v,\label{VP2new}
\end{gather}
\end{subequations}
where $f_e \left( \bm{r}=( x,y ), \bm{v} = ( v_x,v_y),t \right)$ denotes the generic electron PDF and $\bm{E}$ is the self-consistent electric field associated with the local electron number density, $n_e = \int_{\mathbb{R}^2} f_e \, d^2v$. Further details on the mathematical properties of $f_h$ are provided in Sec.~B of the Supplemental Material.

\begin{figure}[t]
\includegraphics[]{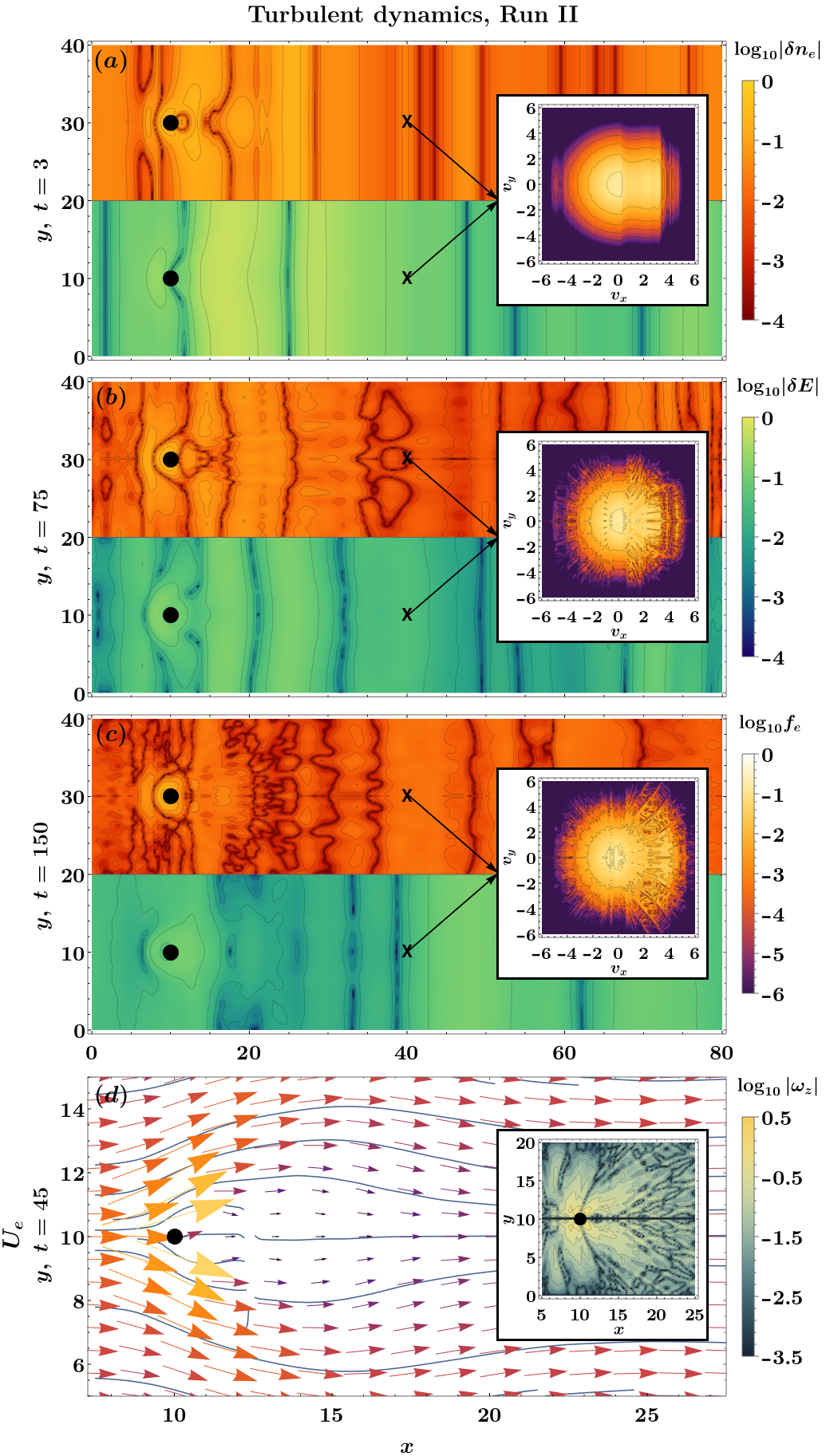}
\caption{Panels (a)--(c): contour plots of the electron density and electric-field fluctuations, $\delta n_e$ and $\delta E$, for Run~II at $t = 3$ [panel (a)], $t = 75$ [panel (b)], and $t = 150$ [panel (c)]. The periodic $y$ domain is shown twice, with $\delta n_e$ and $\delta E$ displayed in the upper and lower halves of each panel, respectively. Insets show the electron PDF in velocity space evaluated at the locations marked by black X symbols. Panel (d): vector plot of the electron bulk velocity $\bm{U}_e$ for Run II at $t = 45$, corresponding to a phase of enhanced vorticity $\omega_z$ (inset). Black disks indicate the positions of the holes.}
\label{fig1}
\end{figure}

Once the equilibrium is determined, its radial profile is used to construct an equilibrium PDF $f_{eq}$ on a discrete 2D-2V phase space, which is then perturbed to study its evolution. The PDF is defined over the periodic spatial domain $\left[ 0, L_x\right] \times \left[ 0, L_y\right]$, while velocity space is restricted to $\left[-L_v,L_v\right]^2$, with the hole placed at $\bm{r}=\left( L_x/8, L_y/2 \right)$. Periodic boundary conditions make the system equivalent to an array of density holes, thereby allowing us to model the wave-hole interaction of $x$-propagating EAWs with repeated kinetic-scale obstacles. The waves are first excited by an external forcing with the appropriate phase velocity \cite{valentini2006excitation,anderegg2009wave,anderegg2009electron,valentini2011new,valentini2012undamped,valentini2025decay,zanelli}, and their fluctuations are then superimposed on $f_{eq}$. The resulting system is advanced in time with a time-splitting VP solver ~\cite{cheng1976integration,filbet2002numerical,valentini2005numerical,valentini2007hybrid,pezzi2013eulerian,celebre2023phase}. More details of the numerical techniques used to solve the VP system and introduce the EAW oscillations are provided in Sec.~A and Sec.~C of the Supplemental Material, respectively.

The aforementioned procedure is applied to four initial conditions defining the numerical runs. Run I is a control simulation without density holes, with an imposed large-scale oscillation $A_0 \sin \left(2\pi y / L_y \right)$ on $n_e$. Run II retains the same setup of Run I, but it includes periodic obstacles with $A_0=0$.  Run III combines obstacles and density perturbation, while Run IV repeats Run III with doubled $L_y$. Further details are provided in Sec.~B of the Supplemental Material.

\textit{Phase-space turbulence}---After constructing $f_{eq}$ and perturbing it with EAWs, the system develops a fully turbulent electrostatic kinetic regime. We find that the presence of density holes strongly affects the turbulent dynamics: the nonlinear cascade initially develops in the $x$--$v_x$ phase subspace, but rapidly spreads to the full four-dimensional phase space.

This behavior shapes the evolution in physical space, as illustrated in Figs.~\ref{fig1}~(a)--(c), which show contour plots of $\delta n_e = n_e - n_{eq}$ and $\delta E = \left| \bm{E} - \bm{E}_{eq} \right|$ for Run II, emphasizing the wave-obstacle interaction over time. Here, $n_{eq}$ and $\bm{E}_{eq}$ stand for the density and electric field associated with $f_{eq}$, respectively. Specifically, the electron deficit modifies the wavefront velocity near the hole, causing waves to slow down around $y\sim L_y/2$ [panel (a)]. This process folds density wavefronts, generating closed structures beyond the holes with small-scale features in both spatial directions [panel (b)]. At later times, waves have interacted repeatedly with the periodic lattice of holes, enhancing the nonlinear dynamics and yielding a turbulent cascade with a complex pattern, especially evident in the $\delta n_e$ distribution [panel (c)].

Additional insight is provided by the vector plot of the electron bulk velocity, $\bm{U}_e = \int_{\mathbb{R}^2} \bm{v} f_e \, d^2v / n_e$. Fig.~\ref{fig1}~(d) illustrates the particle flow around each hole of the array for Run II at $t = 45$, when the vorticity $\omega_z = \left[ \bm{\nabla} \times \bm{U}_e \right]_z$ is enhanced, as shown in the inset. The strongest vorticity is localized near the density depletions. Overall, $\bm{U}_e$ resembles a laminar wake in classical fluid dynamics, with system parameters that inhibit the formation of vortical structures in physical space. Nevertheless, preexisting turbulent structures advected by the mean flow are scattered by the holes, allowing turbulence to spread along both spatial directions.

The development of small-scale structures induced by the inhomogeneous background is not confined to physical space but extends throughout phase space. To illustrate this, we focus on $f_e$ evaluated at $\bm{r} = (L_x/2,L_y/2)$ as a function of $\bm{v}$ [insets of Figs.~\ref{fig1}~(a)--(c)]. At early times, the PDF exhibits smooth contours, displaying a plateau in the positive $v_x$ domain [panel (a)]. The plateau is associated with particles in resonance with EAWs, whose trapping modifies the distribution in such a way as to suppress, according to Landau theory, the damping of the waves themselves \cite{holloway1991undamped,valentini2006excitation,valentini2012undamped}. As time progresses, turbulence generates increasingly complex patterns in velocity space, with both resonant and non-resonant regions being deeply distorted [panels (b)--(c)]. In particular, collimated electron beams with well-defined $v_y/v_x$ ratios emerge.

\begin{figure}[t]
\includegraphics[]{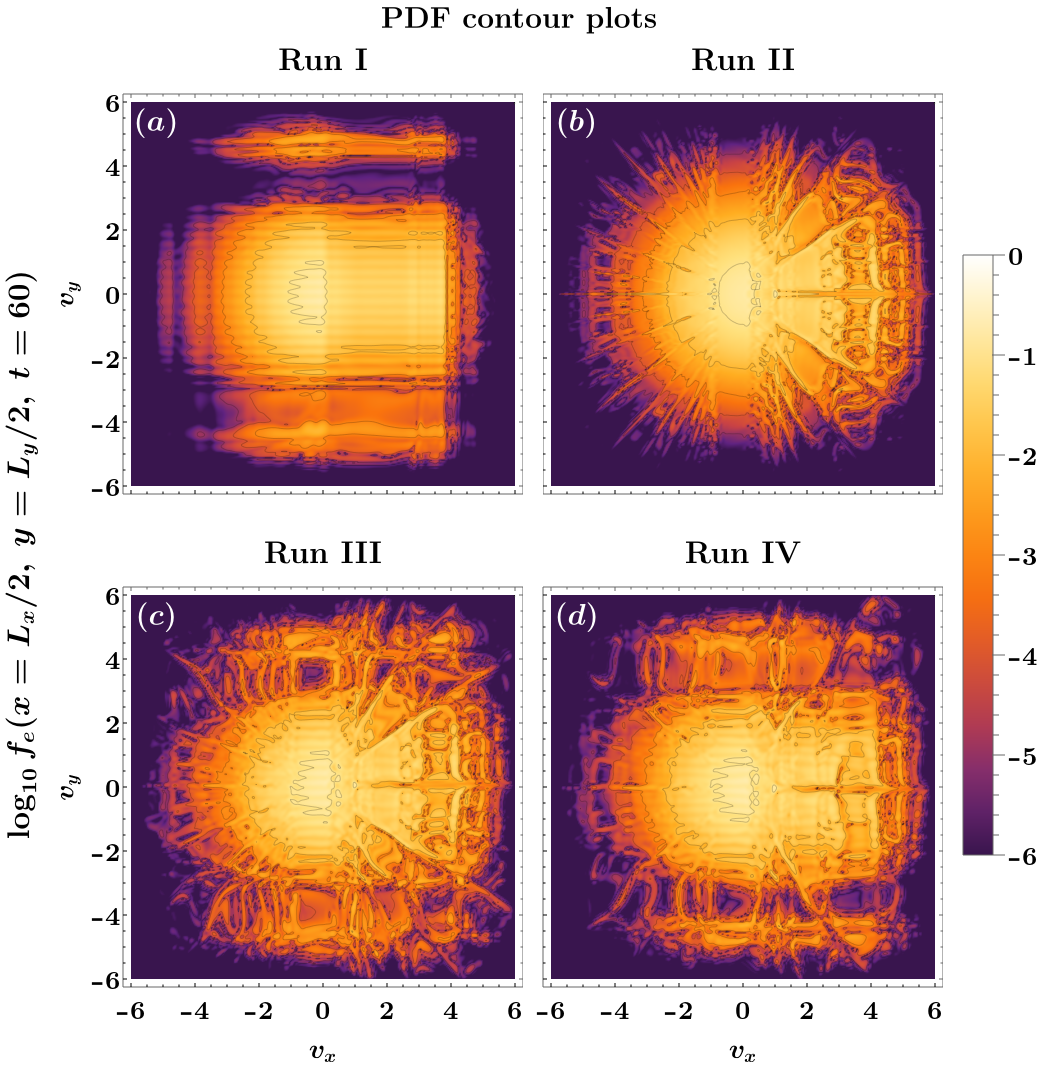}
\caption{Contour plots of the electron PDF in velocity space for each simulation, evaluated at $x = L_x/2$, $y = L_y/2$, $t = 60$.}
\label{fig2}
\end{figure}

To further inspect the nature of these beams, we analyze in $\bm{v}$ space the PDF profile for each simulation at the same fixed position at $t = 60$ (Fig.~\ref{fig2}). The control simulation without holes, i.e., Run I [panel (a)], does not exhibit such beams. In contrast, the $y$ perturbation induces wave-particle interactions at large $v_y$, producing non-thermal tails. This effect can be attributed to Langmuir waves excited by large-scale density fluctuations, as suggested by Run II [panel (b)], where $A_0 = 0$ and phase-space filamentation is absent. When both the hole and the perturbation are present (Run III), all these features coexist, creating a wide variety of small-scale structures [panel (c)]. Furthermore, the outputs from Run IV show that doubling $L_y$ leads to beams with larger opening angles [panel (d)]. This behavior suggests that each hole scatters a fraction of particles in a similar fashion, so that any fixed point receives beams at specific angles determined by the lattice geometry. Consequently, increasing $L_y$ results in a greater vertical spacing and thus a larger minimum arrival angle at that point.

\textit{Spectral analysis}---To quantitatively characterize the turbulent dynamics observed in our framework, we perform a spectral analysis that allows us to isolate the different kinetic scales. This approach provides insight into the cascading mechanisms operating in both spatial and velocity space, in line with previous theoretical studies \cite{dupree1970theory,schekochihin2008gyrokinetic,schekochihin2009astrophysical,schekochihin2016phase,servidio2017magnetospheric,pezzi2019fourier,celebre2023phase,nastac2024phase,nastac2025universal} and consistent with evidence from numerical simulations \cite{celebre2023phase,nastac2024phase,nastac2025universal} and spacecraft measurements \cite{servidio2017magnetospheric}.

An effective method for visualizing PDF fluctuations at each velocity scale is given by the Hermite transform \cite{schumer1998vlasov,parker2015fourier,schekochihin2016phase,servidio2017magnetospheric,adkins2018solvable,pezzi2018velocity,celebre2023phase,zanelli}. This decomposition is based on an orthonormal set of functions whose zeroth mode corresponds to the local Maxwellian distribution, making it particularly useful for isolating the non-thermal component of the velocity distribution, with each mode being associated with a specific moment of the distribution function itself (e.g., Refs.~\cite{schekochihin2016phase,servidio2017magnetospheric,adkins2018solvable,celebre2023phase}). The 2D Hermite transform with index $\bm{m}= (m_x,m_y)$ is defined as:
\begin{align} \nonumber
\hat{f} \left( \bm{r},\bm{m},t \right) = & \int_{\mathbb{R}^2} \bar{f}_e \left( \bm{r},\bm{v},t \right) \\[2pt]
& \times \frac{ H_{m_x} \left(v_x \right) H_{m_y} \left(v_y \right)}{\sqrt{2^{m_x+m_y} m_x! m_y! \pi}} \, e^{-v^2/2} d^2v.
\end{align}

In the expression above, $H_m$ represents the $m$-th Hermite polynomial, while $\bar{f}_e$ is the locally scaled distribution. Denoting by $\bm{T}_e = \int_{\mathbb{R}^2} \left( \bm{v} - \bm{U}_e \right) \otimes \left( \bm{v} - \bm{U}_e \right) \, f_e \, d^2v / n_e$ the electron temperature tensor, and by $U_j^{(e)} (\bm{r},t)$ and $T_{ij}^{(e)} (\bm{r},t)$ the components of $\bm{U}_e$ and $\bm{T}_e$, respectively, the locally scaled distribution satisfies $\bar{f}_e(\bm{r},\bar{\bm{v}},t) = f_e(\bm{r},\bm{v},t)$, with $\bar{v}_j = \left( v_j - U_j^{(e)} \right) / \sqrt{T_{jj}^{(e)}}$.

From Parseval's theorem, the Hermite spectrum determines the spread of fluctuation of the second Casimir invariant in reciprocal velocity space, namely $\Delta C_2$ \cite{knorr1977time,diamond2010modern,servidio2017magnetospheric,nastac2024phase,nastac2025universal}. To perform a meaningful spectral analysis of $\Delta C_2$, which we refer to as enstrophy following Ref.~\cite{servidio2017magnetospheric}, the input data need to be restricted to a homogeneous region. We therefore exclude the hole by considering the subdomain $\mathcal{V}=\left[ L_x/4,3L_x/4\right] \times \left[ 0, L_y \right]$: in this case, $\Delta C_2 = \int_{\mathbb{R}^2} \int_{\mathcal{V}} \Delta f_e^2 \, d^2r \, d^2v / V$, where $\Delta f_e \left( \bm{r},\bm{v},t \right) = \bar{f}_e \left( \bm{r},\bm{v},t \right) - f_{MB} \left( v \right)$, $f_{MB} \left( v \right)=\exp(-v^2/2)/(2\pi)$ is the normalized 2D Maxwellian distribution, and $V = \mu \left( \mathcal{V} \right) = L_xL_y/2$. This quantity plays a central role in phase-space turbulence, as it cascades from the injection range toward dissipative scales in both physical and velocity space, driving the nonlinear turbulent dynamics of kinetic plasmas. Specifically, introducing the local enstrophy as $\Omega (\bm{r},t)$, one has $\Omega = \sum_{\bm{m} \neq \bm{0}} \hat{f}^2$ and $\Delta C_2 = \int_{\mathcal{V}} \Omega \, d^2 r /V$. The spatially-averaged spectrum $F (\bm{m},t)$ is then defined as $F = \int_{\mathcal{V}} \hat{f}^2 \, d^2r /V$.

\begin{figure}[t]
\includegraphics[]{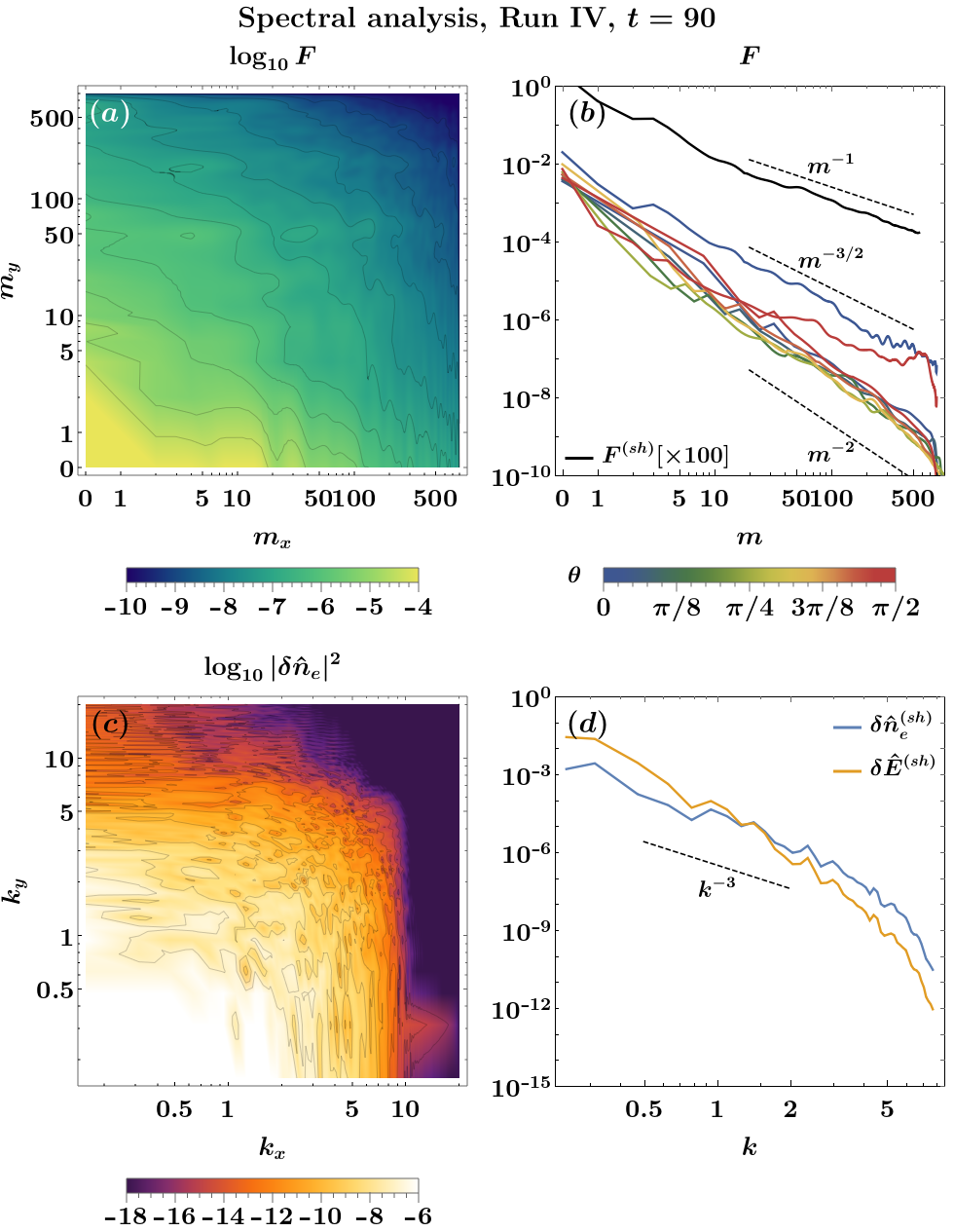}
\caption{Spectral properties of Run IV at $t = 90$. Panel (a): spatially-averaged Hermite spectrum $F(\bm{m})$ in the two-dimensional Hermite space. Panel (b): cuts of the Hermite spectrum in polar coordinates, $F(m,\theta)$, at fixed angles $\theta$, compared with the shell-integrated spectrum $F^{(sh)}(m)$ (black). Panel (c): full Fourier spectrum of density fluctuations $\delta \hat{n}_e(\bm{k})$. Panel (d): shell-integrated Fourier spectra of density and field fluctuations, $\delta \hat{n}_e^{(sh)}(k)$ (blue) and $\delta \hat{E}^{(sh)}(k)$ (orange).}
\label{fig3}
\end{figure}

Fig.~\ref{fig3} displays the Hermite cascade in velocity space and compares it with the Fourier power spectra of $\delta n_e$ and $\delta E$ in physical space for Run IV at $t = 90$. The two-dimensional contour plot of $F(\bm{m})$ for this case [panel (a)] reveals the spread of enstrophy at small scales. The $v_x$ direction remains dominant in the cascade for low Hermite modes, but at intermediate scales ($m \gtrsim 30$) the $\Delta C_2$ flow becomes nearly isotropic.

A more detailed analysis is provided by quantifying the slope of the Hermite spectrum. To this end, we introduce polar coordinates in Hermite space, with radial coordinate $m$ and angle $\theta=\tan^{-1}(m_y/m_x)\in[0,\pi/2]$, and plot cuts of $F$ as a function of $m$ at fixed angles in panel (b). This procedure enables a direct comparison between the numerical results and the cascade models proposed in previous works. Refs.~\cite{servidio2017magnetospheric,pezzi2019fourier} propose a model of velocity-space turbulence in which enstrophy cascades with a constant, isotropic flux down to collisional scales, with a characteristic rate set by the velocity-space advection time, yielding a shell-integrated spectrum scaling as $m^{-3/2}$.

The $m^{-3/2}$ slope has been observed in spacecraft measurements \cite{servidio2017magnetospheric} and in 1D-1V kinetic simulations of wave-driven turbulence \cite{celebre2023phase,zanelli}, and it is also recovered here for $\theta = 0$. The result indicates that a significant fraction of enstrophy cascades predominantly along $v_x$ directly driven by $x$-moving EAWs. This mechanism naturally explains the anisotropy observed in the large-scale contour plots of $F$. In contrast, for $0 < \theta < \pi/2$ and $m \gtrsim 30$, panel (b) indicates that the spectrum is effectively isotropic. In this regime, the spectrum follows a power law $m^{-2}$, corresponding to a shell-integrated scaling as $m^{-1}$. Such an alternative behavior is consistent with the Batchelor-type theory discussed in Refs.~\cite{nastac2024phase,nastac2025universal}, which assumes an enstrophy cascade driven by large-scale $\bm{E}$ fluctuations and sustained by a balance between linear and nonlinear advection in real and velocity space, respectively. Under these assumptions, the aforementioned works derive a reduced spectrum for the Fourier-conjugate variable of $v$, denoted by $s$, which scales as $s^{-1}$. Using the asymptotic relation $s \sim \sqrt{m}$ for $m \gg 1$, this scaling directly translates into an $m^{-1}$ law in Hermite space, since $s^{-1}\,ds \sim m^{-1}\,dm$ in this limit.

The spectral analysis is completed by examining the Fourier transform of the density fluctuations, $\delta \hat{n}_e (\bm{k})$, at $t = 90$. The corresponding spectrum is illustrated in panel (c). Consistent with Fig.~\ref{fig1}, the hole lattice distributes the cascade along both the $x$ and $y$ directions, leading to an approximately isotropic flux. We then inspect, at that time, the density power law through the shell-integrated spectra $\delta \hat{n}_e^{(sh)} (k)$ and $\delta \hat{E}^{(sh)} (k)$ [panel (d)]. At intermediate scales, corresponding to $k \gtrsim 1$, the density spectrum follows a $k^{-3}$ scaling (blue line), again in agreement with Batchelor theory for the 2D-2V VP turbulence model. This corresponds to a $k^{-5}$ scaling for the electric field spectrum (orange line).

The same procedure has also been applied to the scenarios considered in Runs II--III. The corresponding spectra, not shown here, exhibit analogous behavior, suggesting that the spectral scaling and the cascade isotropy for $0<\theta<\pi/2$ are robust properties only weakly affected by the large-scale configuration of the hole lattice.

\begin{figure}[t]
\centering
\includegraphics[]{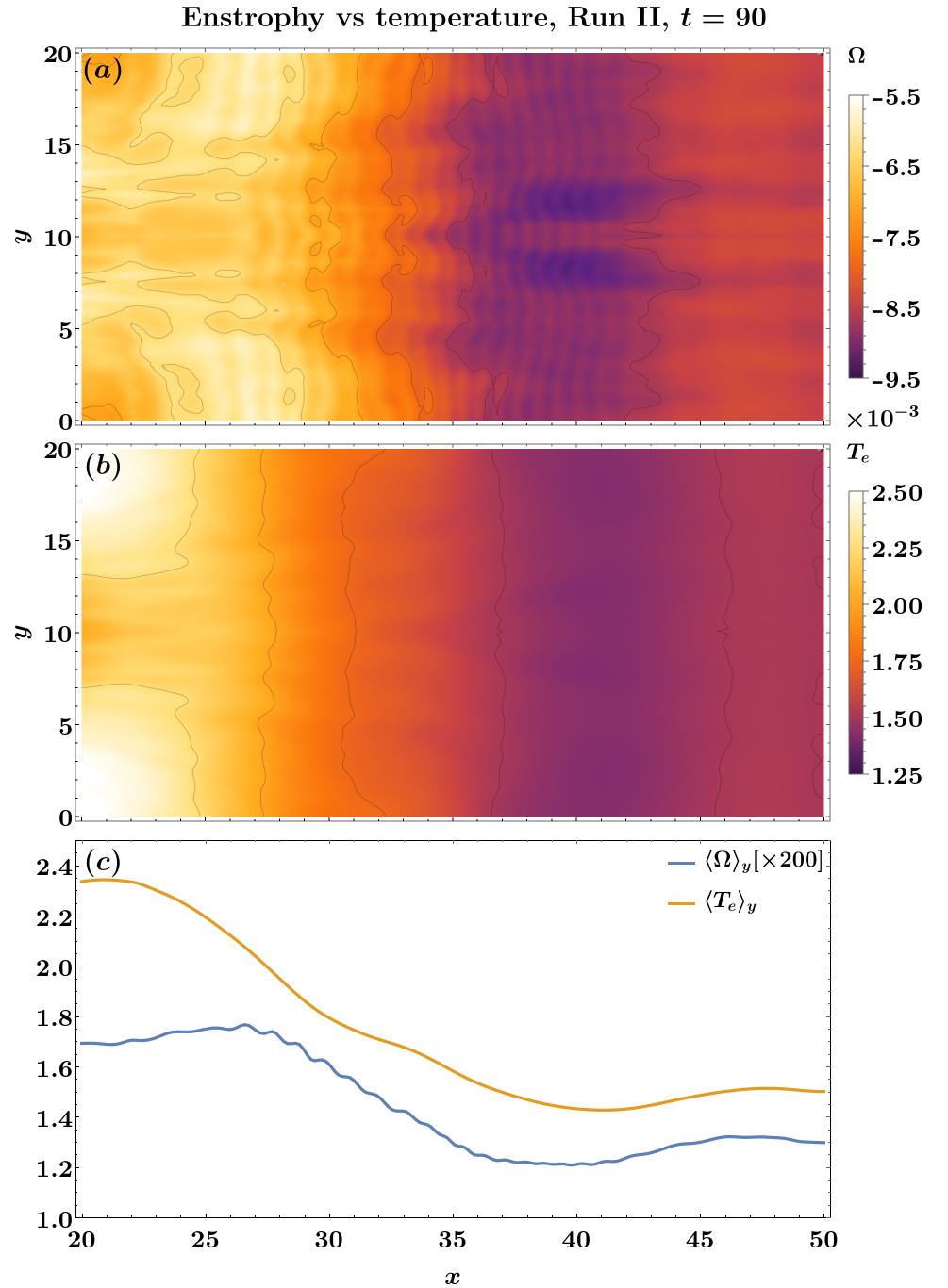}
\caption{Panels (a) and (b): comparison between enstrophy $\Omega$ [panel (a)] and the electron temperature $T_e$ [panel (b)] for Run II at $t=90$. Panel (c): $y$-averaged profiles of enstrophy and electron temperature.}
\label{fig4}
\end{figure}

The development of an enstrophy cascade is widely recognized as a key heating mechanism in weakly collisional plasmas \cite{schekochihin2009astrophysical,pezzi2013eulerian,pezzi2016collisional,pezzi2019fourier,celebre2023phase,nastac2024phase,nastac2025universal}. The spatial locality of this mechanism is not captured by the averaged spectrum $F$. However, the enstrophy density $\Omega$ provides a pointwise measure of the cascade intensity. In particular, $\Omega$ exhibits stronger gradients along $x$ than along $y$, with structures that evolve in time following the propagation of EAWs, while retaining the same overall pattern as the scalar electron temperature
$T_e=\text{Tr}\left(T_{ij}^{(e)}\right)/2$. This behavior is illustrated in Fig.~\ref{fig4}, where $\Omega$ [panel (a)] and $T_e$ [panel (b)] are compared for Run II at $t = 90$. A clearer view is given by the $y$-averaged profiles of these quantities, shown in panel (c). These profiles confirm that regions farther from equilibrium are more strongly heated, as commonly observed in kinetic simulations. A simplified quantitative model describing this correlation is presented in Sec.~D of the Supplemental Material.

\textit{Conclusion}---In this \textit{Letter}, we investigate the behavior of simulated EAWs propagating through a reticulated pattern of stable electron density gaps in 2D–2V phase space. This is the first time that such a configuration has been reproduced and analyzed with the explicit purpose of identifying turbulent cascading mechanisms in kinetic plasma regimes. We find that medium inhomogeneity induces strong distortion of the density and electric-field wavefronts, enhancing wave coupling and promoting the development of turbulence, even orthogonally to the primary wave propagation. In velocity space, the dynamics give rise to complex nonthermal features arising from electron beams scattered by the lattice of density holes, leading to small-scale phase-space structures with a significant role in energy transfer and dissipation at kinetic scales. Our numerical experiments suggest that these structures are associated with a kinetic cascade sustained by the continuous wave-obstacle interaction, pointing to a new paradigm for energy transfer in weakly collisional plasmas.

The electron velocity distribution is analyzed through a Hermite transform, permitting us to identify two main pathways of the kinetic cascade: EAW-driven 1D turbulence and, at intermediate scales, an isotropic 2D enstrophy flux compatible with previous kinetic models of phase-space turbulence. The overall cascade efficiently transfers fluctuations toward dissipative scales, thereby providing a pathway for irreversible plasma heating and entropy production.

These results have direct implications for space plasmas such as the solar wind, where small-scale inhomogeneities can modulate the propagation of Langmuir and other high-frequency waves. Spacecraft measurements can therefore reveal statistical signatures of wave modulation and the associated particle heating, which can be interpreted through high-resolution kinetic simulations.

Future work could explore the role of inhomogeneity parameters, such as the size of density holes and their separation, in setting the turbulent cascade rate and the fluctuation scaling law. Extending the present approach to magnetic kinetic turbulence would provide a deeper insight into space and laboratory plasmas and clarify heating processes under more realistic conditions. Realistic density profiles at Debye scales, in particular, could be directly probed in laboratory experiments, yielding valuable information on the late-stage evolution of plasma turbulence.

\begin{acknowledgments}
The authors are grateful to F. Pegoraro for stimulating discussions on the correlation between temperature and enstrophy. GC and FV acknowledge the support of the PRIN 2022 project ``The ULtimate fate of TuRbulence from space to laboratory plAsmas (ULTRA)'' (2022KL38BK, Master CUP: B53D23004850006), funded by the Italian Ministry of University and Research. SS and FV acknowledge the Space It Up project funded by the Italian Space Agency, ASI, and the Ministry of University and Research, MUR, under contract n. 2024-5-E.0 - CUP n. I53D24000060005. This project has received funding from the European Union's Horizon Europe research and innovation programme under grant agreement No. 101082633 (ASAP).
\end{acknowledgments}

\bibliographystyle{apsrev4-2}
\bibliography{apssamp}

\clearpage
\clearpage
\onecolumngrid

\begin{center}
\vspace*{2em}

\large{\textbf{Supplemental Material for } \it{Plasma Turbulence Driven by Wave-Hole Interaction}}

\vspace{2em}
\end{center}

\twocolumngrid

\preprint{APS/123-QED}

\renewcommand{\thesection}{\Alph{section}}
\setcounter{section}{0}
\renewcommand{\thefigure}{S.\arabic{figure}}
\setcounter{figure}{0}
\renewcommand{\thetable}{S.\arabic{table}}
\setcounter{table}{0}
\renewcommand{\theequation}{S.\arabic{equation}}
\setcounter{equation}{0}

\subsection{Computational Approach to Vlasov-Poisson Dynamics}

The 2D–2V Vlasov–Poisson (VP) system corresponding to the physical configuration analyzed in the main text---namely, an electrostatic equilibrium whose perturbations oscillate near the electron plasma frequency---reads:
\begin{subequations}
\label{VP}
\begin{gather}
\dfrac{\partial f_{e}}{\partial t} + \bm{v} \cdot \bm{\nabla} f_{e} - \dfrac{e}{m_e} \bm{E} \cdot \bm{\nabla}_{\bm{v}} f_e = 0, \label{VP1} \\[5pt] 
\bm{\nabla} \cdot \bm{E} = 4 \pi e \left( n_0 - \int_{\mathbb{R}^2} f_e \, d^2v \right),\label{VP2}
\end{gather}
\end{subequations}
where $f_e$ is the electron particle distribution function (PDF), $\bm{E}$ is the associated electric field, and $n_0$ denotes the uniform density of the stationary proton background. Here, $e$ and $m_e$ are the elementary charge and electron mass, respectively.

Within this notation, a stationary PDF satisfies Eqs.~\eqref{VP1}--\eqref{VP2}, under the condition $\partial /\partial t = 0$, if it depends solely on the single-particle energy $\mathcal{E} = m_ev^2/2-e\psi$, with $\bm{E} = -\boldsymbol{\nabla} \psi$. In this case, a physically consistent choice for $f_e$ is given by:
\begin{equation} \label{fhdim}
f_{e} (\bm{r},v) = \frac{n_0}{2 \pi v_{th, e}^2} \exp{\left( - \frac{v^2}{2 v_{th, e}^2} + \frac{e \psi ( \bm{r})}{m_e v_{th, e}^2} \right)},
\end{equation}
where $v_{th, e} = \sqrt{k_BT_{e, 0}/m_e}$ is the electron thermal velocity associated with the equilibrium electron temperature $T_{e,0}$, and $k_B$ is the Boltzmann constant. The function $f_h$ is chosen to asymptotically recover the Maxwellian distribution $f_{MB}(v)$ when the associated potential $\psi_h$ vanishes.

To properly address the numerical solution of the VP system, all relevant quantities introduced in the main text are recast in normalized units. The normalization relies on the standard characteristic scales of kinetic plasmas, namely the electron Debye length $\lambda_{D,e} = \sqrt{k_B T_{e,0} /(4 \pi n_0 e^2)}$, the electron plasma frequency $\omega_{p,e} = \sqrt{4 \pi n_0 e^2/m_e}$, and the thermal velocity $v_{th, e}$. More precisely, the phase–space domain of $f_e$ is rescaled according to the transformations:
\begin{subequations}
\begin{align} 
&\bm{r} \rightarrow \lambda_{D,e} \bm{r}, \qquad \bm{\nabla} \rightarrow \lambda_{D,e}^{-1} \bm{\nabla}, \\[5pt]
&\bm{v} \rightarrow v_{th,e} \bm{v}, \qquad \bm{\nabla}_{\bm{v}} \rightarrow v_{th,e}^{-1} \bm{\nabla}_{\bm{v}}, \\[5pt] 
& t \rightarrow \omega_{p,e}^{-1} t. \end{align}
\end{subequations}
In this way, Eqs.~\eqref{VP}--\eqref{fhdim} reduce to Eqs.~(1)--(2), provided that the distribution function, electric field, and electrostatic potential are also rescaled as:
\begin{subequations}
\begin{align}
&f_e \rightarrow \frac{n_0}{v_{th,e}^2} f_e, \\[5pt] 
&\bm{E} \rightarrow \frac{k_BT_{e,0}}{e \lambda_{D,e}} \bm{E}, \qquad \psi \rightarrow \frac{k_BT_{e,0}}{e} \psi .
\end{align}
\end{subequations}

The choice of domain parameters and fluctuation properties defines a set of distinct configurations, whose dynamics are computed using a time-splitting scheme that represents the higher-dimensionality extension of the algorithm used by \citet{celebre2023phase}. The method is an advanced, parallelized version of the algorithm originally introduced by \citet{cheng1976integration} (see also  \cite{mangeney2002numerical,filbet2002numerical,valentini2005numerical,valentini2007hybrid,pezzi2013eulerian}), here extended to four dimensions. The underlying idea is to decompose the 1D-1V Vlasov equation for a generic distribution $f$ into separate advection equations:
\begin{align}
&\frac{\partial f}{\partial t} + v_x \frac{\partial f}{\partial x} - E_x \frac{\partial f}{\partial v_x} = 0 \nonumber \\[2pt]
\longrightarrow & \begin{cases}
\dfrac{\partial f_1}{\partial t} + v_x \dfrac{\partial f_1}{\partial x} = 0 \\[5pt]
\dfrac{\partial f_2}{\partial t} - E_x \dfrac{\partial f_2}{\partial v_x} = 0
\end{cases},
\end{align}
whose analytical solutions are $f_1(x,t) = f_1(x - v_x t, 0)$ and $f_2(v_x,t) = f_2(v_x + \int_0^t E_x dt', 0)$, respectively. On a discrete grid in $x$ and $v_x$, the evolution of $f_1$ and $f_2$ can be efficiently approximated using the translation operators $\Lambda_x$ and $\Lambda_{v_x}$:
\begin{subequations}
\begin{gather}
f_1(x,t+\Delta t) \approx \Lambda_x (\Delta t) f_1 (x,t) = f_1(x-v\Delta t,t), \\[5pt] 
f_2(v_x,t+\Delta t) \approx \Lambda_{v_x} (\Delta t) f_2 (v_x,t) = f_2 (v+E_x \Delta t,t).
\end{gather}
\end{subequations}
By adopting the third-order Van Leer scheme for translations \cite{mangeney2002numerical,valentini2005numerical,valentini2007hybrid,pezzi2013eulerian,celebre2023phase}, the operators can be combined to yield the algorithm used by \citet{celebre2023phase}:
\begin{align} \label{trans}
f(x,v_x,t+\Delta t) \approx \, & \Lambda_x\left(\frac{\Delta t}{2}\right) \Lambda_{v_x}(\Delta t)\Lambda_x\left(\frac{\Delta t}{2}\right) f(x,v_x,t) \nonumber \\[2pt]
= \, & \Xi_x(\Delta t) f (x,v_x,t),
\end{align}
where $E_x$ in $\Lambda_{v_x}$ is computed by solving the Poisson equation for the distribution $\Lambda_x (\Delta t/2) f(t)$ through a standard FFT solver. The approximation error associated with Eq.~\eqref{trans} is $\mathcal{O}(\Delta t^3)$ \cite{mangeney2002numerical}.

The extension of the splitting scheme to the electron distribution $f_e$ in 2D-2V phase space reads:
\begin{align}
&\dfrac{\partial f_{e}}{\partial t} + \bm{v} \cdot \bm{\nabla} f_{e} - \bm{E} \cdot \bm{\nabla}_{\bm{v}} f_e = 0 \nonumber \\[2pt]
\longrightarrow & \begin{cases} 
\dfrac{\partial f_x}{\partial t} + v_x \dfrac{\partial f_x}{\partial x} - E_x \dfrac{\partial f_x}{\partial v_x} = 0 \\[5pt]
\dfrac{\partial f_y}{\partial t} + v_y \dfrac{\partial f_y}{\partial y} - E_y \dfrac{\partial f_y}{\partial v_y} = 0
\end{cases},
\end{align}
which, analogously to Eq.~\eqref{trans}, defines the solution procedure employed in this work:
\begin{align}
f_e(\bm{r},\bm{v},t+\Delta t) = \, & \Xi_x\left(\frac{\Delta t}{2}\right) \Xi_y(\Delta t)\Xi_x\left(\frac{\Delta t}{2}\right) f_e( \bm{r},\bm{v},t) \nonumber \\[2pt]
 & + \, \mathcal{O} (\Delta t^3) .
\end{align}

\subsection{Initial Setup and Equilibrium Density Hole Configuration}

In our work, we represent a stationary electron hole through an inhomogeneous equilibrium PDF $f_h(r, v)$, assumed to possess cylindrical symmetry in physical space. This equilibrium has the form of Eq.~\eqref{fhdim} in dimensionless units,
\begin{equation} \label{fh}
f_{h} (r,v) = \frac{1}{2 \pi} \exp{\left( - \frac{v^2}{2} + \psi_{h} (r) \right)},
\end{equation}
with $\psi_h$ corresponding to a radially symmetric depletion in the density field. The associated electric field and potential are $\bm{E}_{h} = E_h(r)\,\hat{\bm{r}}$ and $\psi_{h} = \psi_{h}(r)$, respectively. The field $\bm{E}_{h}$ satisfies Eq.~(1b) of the main text if and only if:
\begin{equation} \label{diveq}
\bm{\nabla} \cdot \bm{E}_{h} = 1 - n_h = 1 - \int_{\mathbb{R}^2} f_{h} \, d^2v = 1 - e^{\psi_{h}},
\end{equation}
which, using $\bm{E}_h = -\bm{\nabla} \psi_h$, leads to the following Poisson equation in cylindrical coordinates: 
\begin{equation} \label{pois}
\frac{1}{r} \frac{d}{d r} \left( r \frac{d \psi_{h}}{d r} \right) = e^{\psi_{h}} - 1.
\end{equation}

The numerical problem, therefore, consists in finding a solution of Eq.~\eqref{pois} such that the electron density $n_h$ approaches unity far from the hole, thereby neutralizing the homogeneous proton background, while it drops near the center. However, numerical solutions of Eq.~\eqref{pois} with finite potential at $r=0$ have been proved to be incompatible with the condition $\psi_h \to 0$ as $r \to +\infty$, which is satisfied only when $\lim_{r \to 0} \psi_{h} = -\infty$, namely when $n_h$ vanishes at $r = 0$. This behavior is inconsistent with an outward electric field, $-d\psi_h/dr>0$, as expected from a positive net charge near the origin. This indicates that, in a purely electrostatic regime, an electron hole cannot remain stable: the density gradient tends to be naturally compensated. Conversely, introducing a negative charge at $r = 0$ prevents the hole from collapsing. Importantly, this modification does not alter the formal expression of the VP system, since the origin must, in any case, be excluded from the domain of $\psi_h$.

We now consider the integration of Eq.~\eqref{diveq} over a cylinder $V_{cyl}$, whose axis is parallel to $z$, with radius $r$ and height $H$. Applying the divergence theorem leads to:
\begin{equation}
2 \pi rHE_{h}(r) = \pi r^2 H - 2\pi H \int_{0}^{r} r' e^{\psi_{h}(r')} \, dr' .
\end{equation}
The inclusion of a negative charge with linear density $\lambda$ along the $z$ axis adds an extra term $-\lambda \int_{V_{cyl}} \delta (r) \, dV$ on the right-hand side of the previous equation, leading to:
\begin{equation}
E_{h}(r) = \frac{r}{2} - \frac{1}{r} \int_{0}^{r} r' e^{\psi_{h}(r')} \, dr' - \frac{C}{r},
\end{equation}
where $C = \lambda / (2 \pi)$. By defining $I(r) = \int_0^r r' e^{\psi_h(r')} \, dr'$, we obtain a new system of equations to be solved numerically:
\begin{subequations}
\label{system}
\begin{gather}
\frac{d\psi_h}{dr} = - \frac{r}{2} + \frac{I}{r} + \frac{C}{r}, \label{systema} \\[5pt]
\frac{dI}{dr} = r e^{\psi_h}. \label{systemb}
\end{gather}
\end{subequations}
The additional term $C/r$ in the above expression controls the behavior of $-d\psi_h/dr<0$ as $r\to 0$, physically justifying each potential profile corresponding to a hole, whose shape is determined by $C$. On the other hand, each solution of Eqs.~\eqref{systema}--\eqref{systemb} also satisfies Eq.~\eqref{fh} for $r>0$, where $\delta(r)$ vanishes. In this case, we can calculate the potential profile in a numerical domain starting from $r_{min}$ very close to the origin, using $\psi_h(r_{min})$ and $C$ as parameters; $I(r_{min})$ is negligible and therefore fixed at zero. Setting $r_{min} = 10^{-3}$, $\psi_h (r_{min}) = -20$ and $C = 2.66799905$ in a fourth-order Runge-Kutta scheme, we get a desirable profile for both the potential and the electron density: $\psi \left( r>r_h=10\right)$ is negligible, allowing us to define $r_h$ as the hole radius.

\begin{table}[t]
\caption{\label{tab1}%
Parameters of the simulations performed. In all runs: $L_x=80$, $N_x=512$, $L_v=6$, $N_v=301$, $\Delta t=2 \times 10^{-3}$, and $r_h = 10$.}
\begin{ruledtabular}
\begin{tabular}{ccccc}
Run&$f_{eq}$&$A_0$&$L_y$&$N_y$\\
\hline
I&$f_{MB}$&0.2&20&128 \\
II&$f_{unp}(t_1)$&0&20&128 \\
III&$f_{unp}(t_1)$&0.2&20&128 \\
IV&$f_{unp}(t_1)$&0.2&40&256 \\
\end{tabular}
\end{ruledtabular}
\end{table}

\begin{figure*}[!t] 
\centering
\includegraphics[]{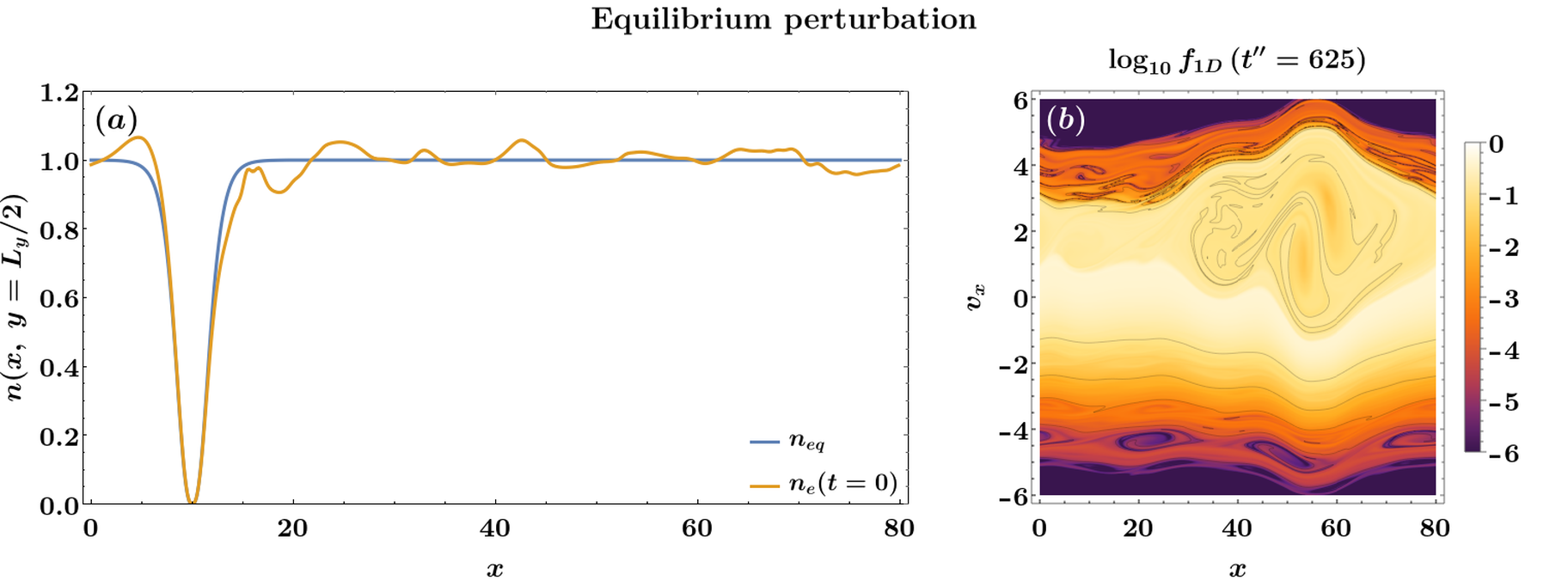}
\caption{\label{fig:app1} Panel (a): inhomogeneous electron density at equilibrium (blue) and after perturbation (orange), evaluated at $y = L_y/2$. Panel (b): contour plot of the 1D–1V distribution function $f_{1\mathrm{D}}(t_2)$, associated with right-propagating turbulent EAWs and used to construct $\delta f_e(t=0)$.}
\end{figure*}

Once $f_h$ is determined, its expression is used to construct the equilibrium electron distribution $f_{eq}$ for a set of four simulations (Runs I--IV) on a four-dimensional uniform grid with $N_x\times N_y\times N_v^2$ mesh points. Run I is a control simulation without any hole, in which charge neutrality is preserved at equilibrium, and the electrons are distributed according to $f_{MB}$. In the remaining simulations, $f_{eq}$ is built from an unperturbed distribution $f_{unp}$, obtained by embedding the radial hole profile into the spatial domain $\left[ 0, L_x\right] \times \left[ 0, L_y\right]$ as:
\begin{equation}
f_{unp} (\bm{r},v,t'=0) = \begin{cases} 
f_{h} (r',v) \;\;\; \text{if } r' \leq r_h \\
f_{MB} (v) \;\;\;\; \text{if } r' > r_h
\end{cases},
\end{equation}
where $r' = \sqrt{\left( x -L_x/8\right)^2+\left( y -L_y/2\right)^2}$. The domain sizes $L_x$ and $L_y$ are chosen such that the excited modes along the $x$ direction have wavelengths comparable to $r_h$, but much smaller than the $x$-spacing between consecutive holes. This choice enables us to probe the wave dynamics both near the density depletion and far from it. The unperturbed distribution $f_{unp}$ then evolves over an auxiliary time $t'$ through the time-splitting scheme to assess the stability of this state. We observe that the numerical error, $\max \lbrace \left| f_{unp}(t') - f_{unp} (0) \right|\rbrace$, grows during the first plasma periods and then saturates at approximately $8 \times 10^{-5}$ around $t'=t_1=15$. For this reason, we assume $f_{eq}=f_{unp} (t_1)$ in Runs II--IV, which are distinguished by the arrangement of density holes and the initial transverse modulation. In Run II, $f_{eq}$ remains unperturbed along the $y$ direction. In contrast, in Run III an oscillation, $A_0 \sin \left(k_{0y} \, y \right)$, is initially added to the electron density $n_e$, with $A_0 = 0.2$ and $k_{0y} = 2 \pi/L_y$. Run IV adopts the same initial conditions as Run III, except that $L_y$ is doubled. The characteristics of each simulation are sketched in Table~\ref{tab1}.

\begin{figure}[t]
\centering
\includegraphics[]{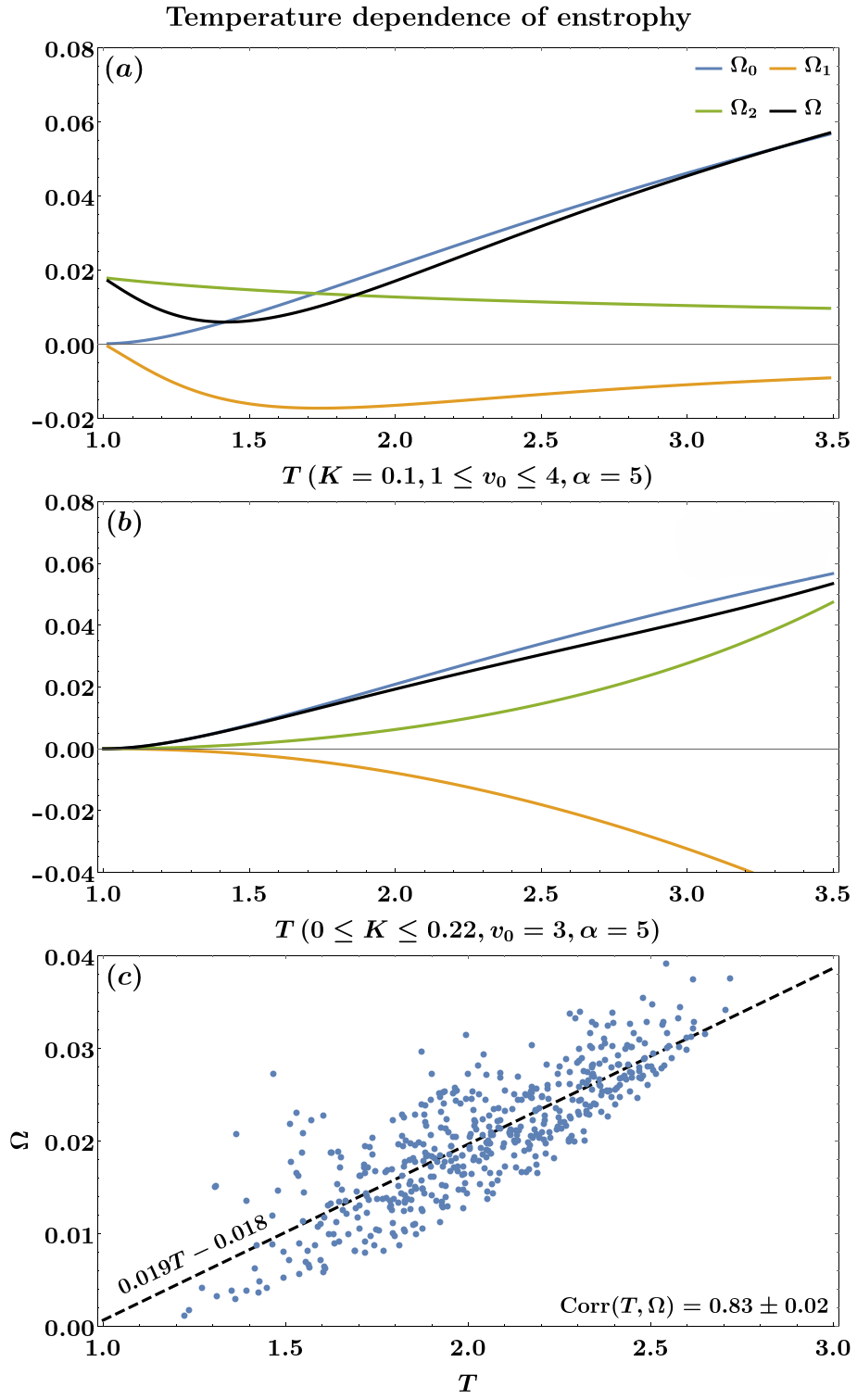}
\caption{\label{fig:app2}Panels (a) and (b): enstrophy $\Omega \left( K, v_0, \alpha \right)$ in the single beam-pair model (black line), together with its components $\Omega_0$, $\Omega_1$, and $\Omega_2$, as functions of the electron temperature $T \left( K, v_0, \alpha \right)$, for $\alpha = 5$. In panel (a), $K = 0.1$ and $1 \leq v_0 \leq 4$, whereas in panel (b), $v_0 = 3$ and $0 \leq K \leq 0.22$. Panel (c): scatter plot showing the correlation between $T$ and $\Omega$ for multiple beam pairs with random amplitudes.}
\end{figure}

\subsection{Equilibrium Perturbation via Electron Acoustic Waves}

To better assess the role of the density hole in the propagation of plasma waves, it is desirable to initialize the system with an equilibrium perturbation, $\delta f_e$, such that the wave damping is minimized. To this end, we first evolve an auxiliary one-dimensional system, $f_{1D}$, in which oscillations are driven by an external electric field, $\bm{E}_{ext}^{(1D)}$. When the driver matches the correct phase velocity, electron acoustic waves (EAWs) can be excited without significant dissipation, since Landau damping is suppressed by the formation of a plateau in the $v_x$ direction associated with the waves. This process rapidly amplifies the oscillations and drives the system into a highly nonlinear regime. Once this turbulent state is established, the resulting deviation from thermal equilibrium is consistently adopted as the initial perturbation for the two-dimensional configuration.

The 1D simulation is performed over the domain $[0, L_x] \times [-L_v, L_v]$, discretized with a high-resolution grid of $N_x \times N_v'$ points ($N_v' = 3001$). Following Refs.~\cite{valentini2006excitation,valentini2012undamped,valentini2013response,valentini2025decay,zanelli}, the system is initialized in thermal equilibrium, represented by the 1D Maxwellian $f_{MB}^{(1D)}(v_x) = \exp(-v_x^2/2) / \sqrt{2\pi}$. The distribution is then evolved over an auxiliary time $t''$ under the action of $\bm{E}_{ext}^{(1D)}=E_{ext}^{(1D)} \hat{\bm{x}}$, given by:
\begin{align} \label{Eext} \nonumber
E_{ext}^{(1D)}(x, t'') =& \, E_0 \left[1 + \left(\frac{t'' - \tau_0}{\delta \tau}\right)^{\beta}\right]^{-1} \\[2pt]
& \times \sum_{n=2}^{4} \frac{1}{n} \sin\left(k_n x - \omega_n t'' + \varphi_n\right), 
\end{align}
where $k_n = n k_{0x} = 2\pi n/L_x$ denotes the $n$-th excited $x$ wavenumber, $\omega_n = v_\varphi k_n$ the corresponding frequencies (with $v_{\varphi}$ the phase velocity), $E_0 = 0.25$ the driver amplitude, and $\varphi_n$ randomly assigned phases. The parameters $\tau_0 = -100$, $\delta \tau = 500$, and $\beta = 40$ prescribe a smooth temporal window to activate the external field. The intensity of $E_{ext}$ is approximately $E_0$ at the beginning of the simulation, decreases to $E_0/2$ at $t'' = 400$, and becomes negligible after $t'' \simeq 500$. The phase velocity is set to $v_\varphi = 2$ for each excited wavelength. In this way, the forcing excites plasma modes lying close to the thumb curve in $k$--$v_\varphi$ space (see Refs.~\cite{holloway1991undamped,valentini2006excitation,valentini2011new,valentini2012undamped,valentini2013response} for details), thereby ensuring, within the framework of quasi-linear theory, the rapid growth of EAW-like modes. These modes overcome Landau damping and drive the system into a strongly turbulent state after several trapping times. Specifically, we observe the emergence of high-amplitude, $x$-propagating plasma waves at several wavenumbers $k_x$, with phase speed $\sim v_{\varphi}$. This behavior arises from nonlinear wave coupling, which manifests in the 1D-1V phase space as right-moving vortices. These vortices induce a direct cascade that transports fluctuations toward smaller scales, while simultaneously promoting an inverse cascade via the coalescence of large-scale structures.

Once a scenario is achieved where all these phenomena occur simultaneously, at $t'' = t_2 = 625$, the distribution function $f_{1D}$ is rescaled to the coarse-grained grid and adapted for the 2D simulation with the density hole:    
\begin{equation}
f_e(t = 0) = f_{eq} + \delta f_e(t = 0) = f_{eq} + W[f_{1D}(t_2)],
\end{equation}
where
\begin{align} \nonumber
W[f(x, v_x)] = & \, w(x) \Big(A f(x - x^\ast, v_x) \\[2pt] 
&  - f_{1D}^{(MB)}(v_x) + B\Big) \, f_{1D}^{(MB)}(v_y).
\end{align}
Here, $w(x)$ is a window function that preserves the shape of the density hole and, for instance, prevents the appearance of negative densities. The Gaussian factor defines the Maxwell-Boltzmann profile in the $v_y$ direction, which remains unaffected by the 1D turbulence. The parameters $A$, $B$, and $x^\ast$ are chosen such that $f_e$ and $f_{eq}$ contain the same total mass. This condition is enforced by requiring that the (numerical) integral of $W[f_{1D}]$ over the full phase space vanishes:
\begin{align} \label{Wint}
\!\!\!\!&\int_{\mathbb{R}^4} W[f_{1D}(t_2)] \, d^2r \, d^2v \nonumber \\[2pt] 
\!\!\!\!=& L_y \int_0^{L_x} w(x)\,
    \Big( A n_{1D}(x - x^\ast) - 1 + B' \Big) \, dx = 0 .
\end{align}
where $n_{1D} = \int f_{1D}(t_2) \, dv_x$ and $B' = \int B \, dv_x = (2N_v - 1) \left( L_v/N_v \right) B$. The relation between $B$ and $B'$ follows directly from the use of the rectangle rule in the integration. The resulting $n_e(t = 0)$ and $f_{1\mathrm{D}}(t_2)$ are shown in Fig.~\ref{fig:app1}.

To understand the correct choice of the parameters in the equation above, it is useful to first consider the limiting case $A = w = 1$ and $B' = x^\ast = 0$. In this configuration, if $n_{1D}$ is unitary on average, mass is conserved. However, due to the numerical error inherent in the 1D simulation, this condition is not exactly satisfied. We therefore set $A = L_x / \int n_{1D} \, dx \simeq 1.000166$ to compensate for this mismatch. The introduction of a windowing is paramount, as discussed previously. A suitable choice for $w(x)$ is a function that tends to zero inside the hole and to unity far from it. This requirement is fulfilled by selecting $w(x) = n_{h}(x, L_y / 2)$, i.e., the radial profile of the hole. However, the modulation of $\delta n_{1D} = A n_{1D} - 1$ with $w(x)$ does not preserve the integral. For this reason, it is necessary to include the shift $x^\ast$ so that the convolution $w \ast \delta n_{1D} (x)$ vanishes at $x = x^\ast$. A numerical evaluation of the convolution yields $x^\ast \simeq 39.974$ as a suitable value. We then compute a new 1D simulation using the transformation $x \rightarrow x - x^\ast$ in Eq.~\eqref{Eext}. Grid effects nevertheless prevent Eq.~\eqref{Wint} from being satisfied down to round-off error unless $B'$ is assigned a small non-zero value. In particular, we find $B' \simeq 3.651 \times 10^{-7}$, and consequently $B \simeq 3.032 \times 10^{-8}$.

\subsection{Enstrophy-Temperature Correlation}

Fig.~4 of the main text suggests a possible spatial correlation between the electron kinetic temperature and the enstrophy budget. This is qualitatively consistent with the role of the enstrophy cascade in plasma heating: strong departures from a Maxwellian distribution are expected to be associated with an enhanced tendency of the system to dissipate fluctuations and restore thermal equilibrium through an effective small-scale collisionality associated with grid-scale effects in our simulations. On the other hand, the relationship between $\Omega$ and $T_e$ is nontrivial, since they are obtained by integrating over velocity space different quantities associated with $f_e$, namely $\Delta f_e^2$ and $v^2 f_e$, respectively.

This relationship can be quantitatively clarified by considering a simplified model distribution function defined in one-dimensional velocity space. As a reference equilibrium distribution, we take the Maxwellian $f_{1D}^{(MB)}$ with unit temperature. We then consider a perturbation consisting of a pair of symmetric beams, $\delta f_{\pm}=\sqrt{\alpha/\pi}\,K
\exp\left[ -\alpha \left( v \mp v_0 \right)^2 \right]$, centered at $v = \pm v_0$ and characterized by a thermal width $(2\alpha)^{-1} \ll 1$. These peaks affect both temperature $T$ and enstrophy $\Omega$ of $f = f_{1D}^{(MB)} + \delta f_{+} + \delta f_{-}$ in a controlled way. Since the bulk velocity of $f$ vanishes by symmetry, one has
\begin{equation}
T = \frac{\int_{-\infty}^{+\infty} v^2 f \, dv}{\int_{-\infty}^{+\infty} f \, dv} = \frac{1 +K \left( 2v_0^2 + \alpha^{-1} \right)}{1 + 2K} . \label{temp}
\end{equation}
In our simulations, the temperature increases globally in time. To reproduce this feature within the model, we choose beams located sufficiently far from the Maxwellian core, $v_0 \gtrsim 1$, so that the resulting distribution has a temperature larger than that of the reference Maxwellian, namely $T > 1$.

To compute enstrophy, we first express the distribution in the normalized velocity coordinate, namely $\bar{f}(v)=f(\sqrt{T}v)$, which gives:
\begin{align}
\Omega = & \int_{-\infty}^{+\infty} \left( \bar f- f_{1D}^{(MB)}\right)^2 \, dv \nonumber \\[2pt] 
= & \int_{-\infty}^{+\infty} \bigg[ \frac{1}{\sqrt{2\pi}} \left(e^{-Tv^2/2} - e^{-v^2/2} \right) \nonumber \\[2pt] 
& + \, \sqrt{\frac{\alpha}{\pi}} K \sum_{\mp} e^{-\alpha \left( \sqrt{T} v \mp v_0 \right)^2}  \bigg]^2 \, dv \nonumber \\[2pt]
=& \int_{-\infty}^{+\infty} \left( G_T + G_0 + G_+ + G_- \right)^2 \, dv . \label{enstrophy}
\end{align}
The integral can be written as $\Omega = \sum_{i,j} I_{ij}$, with $I_{ij} = \int_{-\infty}^{+\infty} G_i G_j \, dv$. After straightforward algebra, enstrophy can be decomposed into three contributions, $\Omega = \Omega_0 + \Omega_1 + \Omega_2$. The first one, $\Omega_0 = \sum_{i,j\in\{T,0\}} I_{ij}$, is independent of the beam amplitude $K$. The second one, $\Omega_1 = 2\sum_{i\in\{T,0\}} \sum_{j\in\{+,-\}} I_{ij}$, is linear in $K$, while $\Omega_2 = \sum_{i,j\in\{+,-\}}I_{ij}$ is quadratic in $K$. Their explicit expressions are:
\begin{subequations}
\begin{align}
\!\!\!\! \Omega_0 &= \frac{1}{2 \sqrt{\pi}} \left( 1 + \frac{1}{\sqrt{T}} - \frac{2\sqrt{2}}{\sqrt{1+T}}\right), \label{omega0}\\[5pt]
\!\!\!\! \Omega_1 &= 4 \sqrt{\frac{\alpha}{\pi}} \left[ \frac{e^{- \alpha v_0^2/\left( 2 \alpha + 1 \right)}}{\sqrt{\left( 2 \alpha + 1 \right) T}} - \frac{e^{- \alpha v_0^2/\left( 2 \alpha T + 1 \right)}}{\sqrt{\left( 2 \alpha T + 1 \right)}} \right] K, \label{omega1} \\[5pt]
\!\!\!\! \Omega_2 &\approx \sqrt{\frac{2 \alpha}{\pi T}} K^2. \label{omega2}
\end{align}
\end{subequations}
Here, the mixed beam-beam contributions $I_{+-}$ and $I_{-+}$ are equal by symmetry and have been neglected in $\Omega_2$, since they are exponentially small for well-separated beams because $\alpha v_0^2\gg 1$.

Even in this simplified model, the relationship between $\Omega$ and $T$ is nontrivial. Nevertheless, the dependence $\Omega(T)$ exhibits clear trends. As an illustrative case, panel~(a) of Fig.~\ref{fig:app2} shows $\Omega_0$, $\Omega_1$, and $\Omega_2$ for temperatures in the range $1 \leq T \leq 3.5$, chosen to be comparable with those observed in our simulations. Since both $T$ and enstrophy contributions depend on $K$, $v_0$, and $\alpha$, we fix $\alpha = 5$ and $K = 0.1$ in order to represent how the enstrophy components vary with temperature. With these values, $T$ spans the range of interest when the remaining free parameter satisfies $1 \leq v_0 \leq 4$, according to Eq.~\eqref{temp}. The total enstrophy is represented by a black line.

For $T \gtrsim 1.5$, $\Omega_0$ increases almost linearly with $T$. In this temperature range, the expression in Eq.~\eqref{omega0} is well approximated by the linear term of its local expansion, while higher-order contributions remain subdominant. By contrast, both $\Omega_1$ and $\Omega_2$ decrease as $v_0$ and, consequently, $T(v_0)$ increase, in agreement with Eqs.~\eqref{omega1}--\eqref{omega2}. Therefore, at sufficiently high temperatures, total enstrophy is dominated by $\Omega_0$, yielding $\Omega \approx \Omega_0$ and an approximately linear dependence on $T$.

A similar behavior is recovered in panel~(b) of Fig.~\ref{fig:app2}, where we keep the same value of $\alpha$, set $v_0 = 3$, and vary the beam amplitude in the range $0 \leq K \leq 0.22$, obtaining the same temperature interval as in panel~(a). In this case, $\Omega_1$ and $\Omega_2$ increase with $T(K)$ but have opposite signs, so that their combined contribution to $\Omega$ remains nearly negligible. This compensation no longer holds for larger values of $K$, since $\Omega_1 \propto K$ whereas $\Omega_2 \propto K^2$; however, this occurs at temperatures outside the range explored in our investigation. Finally, we do not explore the dependence on $\alpha$ in detail, since for $\alpha \gg 1$ its contribution to $T$ in Eq.~\eqref{temp} is small. Nevertheless, we verified that $\Omega(T)$ preserves the approximately linear trend shown in panels~(a) and~(b) for $\alpha \lesssim 20$.

So far, we have considered a simplified case in which thermal equilibrium is perturbed by a single pair of peaks. A more realistic scenario can be obtained by modeling the velocity-space oscillations through multiple Gaussian beams, leading to expressions for $T$ and $\Omega$ analogous to Eqs.~\eqref{temp}--\eqref{enstrophy}. In particular, the same expression for $\Omega_0$ as in Eq.~\eqref{omega0} is retained, since this term does not explicitly depend on the shape of the perturbation. On the other hand, $\Omega_1$ and $\Omega_2$ are expected to be given by sums of contributions analogous to Eqs.~\eqref{omega1}--\eqref{omega2}, weighted by the amplitudes and locations of the different peaks, including possible cross terms. Therefore, this model suggests that a broad class of small-scale perturbations can lead to correlated variations of enstrophy and temperature.

To verify this, we numerically compute $T$ and $\Omega$ from a sample of 500 distribution functions constructed by adding four pairs of symmetric beams with centers $\lbrace v_b \rbrace$ satisfying $|v_b|>1$ and random amplitudes in the range $0 \leq K_\ell \leq 0.08$. These beams mimic the strong velocity-space oscillations observed in the distribution functions from our simulations. The resulting configurations are scattered in the $T$-$\Omega$ plane, as shown in panel~(c) of Fig.~\ref{fig:app2}, and exhibit a strong correlation between the two quantities. The data are well described by the linear fit $\Omega \approx 0.019 T - 0.018$, consistent with the condition of negligible enstrophy when $T = 1$. Using a bootstrap procedure, we estimate the correlation coefficient and its standard error, obtaining $\text{Corr}(T, \Omega) = 0.83 \pm 0.02$. This supports the existence of a significant relationship between small-scale velocity-space structures and plasma heating.


%

\end{document}